# Comparison of SMC and OMC results in determining the ground-state and meta-stable states solutions for UO$_2$ in DFT+U method

M. Payami

*School of Physics & Accelerators, Nuclear Science and Technology Research Institute, AEOI,*

*P. O. Box 14395-836, Tehran, Iran*

**Abstract**

Correct prediction of the behavior of UO$_2$ crystal, which is an antiferromagnetic system with strongly-correlated electrons, is possible by using a modified density functional theory, the DFT+U method. In the context of DFT+U, the energy of crystal turns out to be a function with several local minima, the so-called meta-stable states, and the lowest energy state amongst them is identified as the ground state. OMC was a method that were used in DFT+U to determine the ground state. Recently the SMC method was proposed which using only the oxygen electronic spin-polarization degrees of freedom also revealed the multi-minima structure of energy in the DFT+U approach and led to results in good agreement with experiment. In this work, we compare the SMC and OMC results and show that although the ground states of the two methods have similar energies and geometries, the electronic structures have significant differences. Moreover, we show that the GS obtained from SMC is by 0.0022 Ry/(formula unit) above that of OMC. The different GS results from the two methods implies that they search for the minimum-energy state over different subspaces of electron densities and each method alone is not capable to locate the global minimum of the energy. Therefore, to obtain the global-minimum state of energy one has to search over larger subspaces that involve both occupation matrices of U atoms and starting magnetization of O atoms.

## 1. Introduction

Uranium dioxide is a common fuel used in nuclear power reactors. It has a 3k-order anti-ferromagnetic (AFM) crystal structure at temperatures less than 30 K and para-magnetic structure at higher temperatures [1-2]. From earlier experimental results [3] it was concluded that the uranium and oxygen atoms occupied respectively the octahedral (4a) and tetrahedral (8c) symmetry positions of cubic space group $Fm\overline{3}m$ (No. 225) with lattice constant of 5.47 Angstrom, which is shown in Fig. 1(a). This space group can be represented by a simple tetragonal unit cell with 6 atoms as shown in Fig. 1(b).

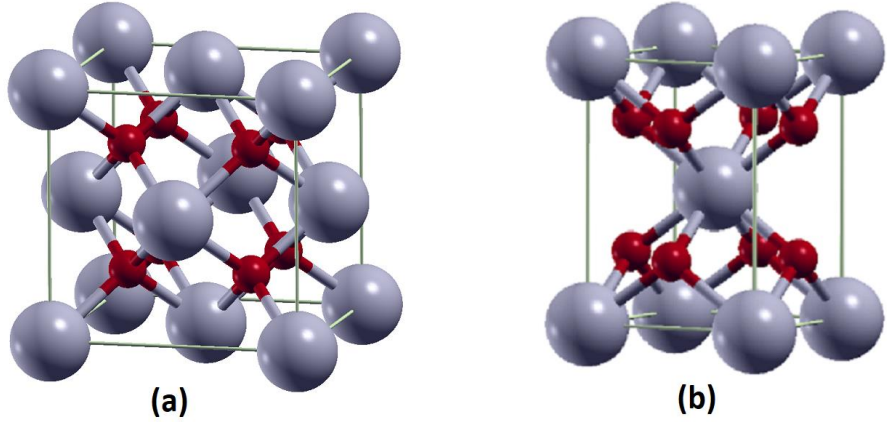

(a)                    (b)

**Fig. 1**. (a)- UO$_2$ crystal structure with cubic space group $Fm\overline{3}m$ (No. 225) and lattice constant of 5.47 Angstrom; (b)- description by a simple tetragonal crystal structure with six atoms. Gray and small red balls represent uranium and oxygen atoms, respectively.

However, a later XRD experiment [4] indicated that UO$_2$ crystallizes with a less symmetric cubic space group $Pa\overline{3}$ (No. 205) with oxygen atoms slightly displaced inside the cube. The use of ordinary approximations in density-functional theory (DFT) [5-6] description of the system leads to incorrect metallic behavior while it is experimentally found [7] to be an insulator with a gap of 2.10 eV. The incorrect metallic prediction arises from the partially-filled "localized" *5f* or *6d* valence electrons in uranium atoms that are treated in the same footing as other delocalized ones in the atom. One of the ways to overcome this problem is the method of DFT+U [8-11] in which the DFT contributions of Hubbard orbitals ($E_{dc}$), are subtracted from the DFT energy and a better contribution, which is borrowed from Hubbard model ($E_{Hub}$), is added to DFT energy, leading to the DFT+U energy functional:

$$E_{DFT+U} = E_{DFT} - E_{dc} + E_{Hub}. \qquad (1)$$

In Eq. (1), the first term in the right hand side is the LDA/GGA [6, 12] total energy of the system. $E_{Hub}$ depends on the elements of the occupation matrices of highly correlated *f* and *d* electrons of uranium atoms. It turns out that $E_{DFT+U}$ is a functional having several local minima, the so-called meta-stable states. To determine the ground state (GS) of the system, one has to carefully determine all such local minimum states and then identify the lowest energy state as the GS of the system. Two of the methods that determine the meta-stable states are called the occupation-matrix control [10] (OMC) and the starting magnetization control [13] (SMC). In this work, we briefly describe these methods and by comparing the results primarily show that the number of metastable states depends on the method and the computational details. Then, it is shown that even though the GS states determined by these two methods have similar energies, the electronic structure of the two GS's have significant differences. Moreover, we have shown that the SMC-GS is by 0.0022 Ry/(formula unit) above the OMC-GS. This difference implies that the two methods search for the minimum-energy state over different subspaces of electron densities and each method alone is not capable to explore the global minimum of the energy. Therefore, to locate the global-minimum state of energy we have searched over a larger subspace which involves at the same time both occupation matrices of Hubbard *f* orbitals of U atoms and the starting magnetization of O atoms and shown that the number of metastable states varies with starting magnetization.

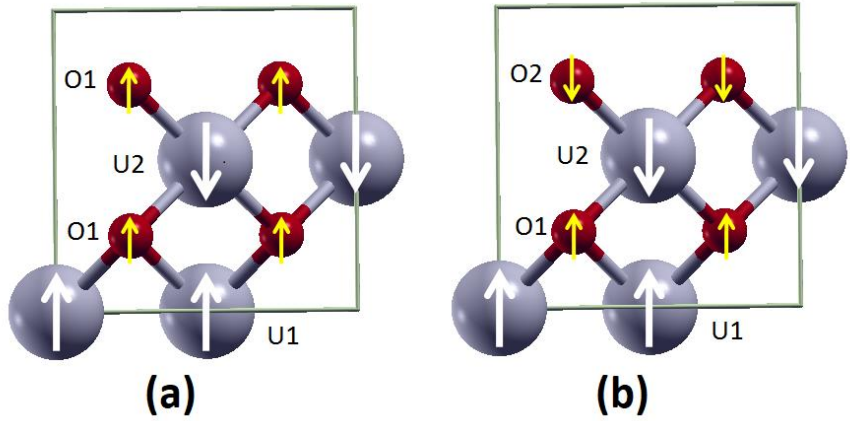

**Fig. 2**. (a)- All-equivalent oxygen atoms scheme; and (b)-two-inequivalent oxygen atoms scheme. In the all-equivalent oxygen atoms model, all oxygen atoms are treated as the same type, O1, and assume the same starting magnetizations within the SMC method; while in two-inequivalent oxygen atoms model, the O1 and O2 oxygen atoms are treated as different types and assume independent values for the starting magnetizations in the SMC method [13].

The organization of this paper is as follows. In Section 2, we present the computational details; in Section 3 the calculated results are presented and discussed; Section 4 concludes this work.

## 2. Computational Details

All DFT+U calculations are based on the solution of the KS equations using the Quantum-ESPRESSO code package [14, 15]. For the atoms U and O, we have employed the ultra-soft pseudo-potentials (USPP) generated by the *atomic* code, using the generation inputs (with small modifications for more desired results) from the *pslibrary* [16]. For the USPP generation, we have used the valence configurations of U($6s2$ $6p6$, $7s2$, $7p0$, $6d1$, $5f3$) and O($2s2$, $2p4$); and to take into account the dominant relativistic effects of the electrons [17], we have adopted the scalar-relativistic (SR) method [18]. For the exchange-correlation, we have chosen the Perdew-Zunger [19] (PZ) LDA, which gives excellent geometric properties whenever the Hubbard on-site parameter $U$ is set to 4.53 eV and the projection on to Hubbard orbitals are chosen to be atomic ones that are not orthogonalized. Performing convergency tests, the appropriate kinetic energy cutoffs for the plane-wave expansions were chosen as 90 and 720 Ry for the wavefunctions and densities, respectively. To avoid the self-consistency problems, the Methfessel-Paxton smearing method [20] for the occupations with a width of 0.01 Ry is used. For the Brillouin-zone integrations in geometry optimizations, a $8\times8\times6$ grid were used; while for density-of-states (DOS) calculations, we have used a denser grid of $10\times10\times7$ in reciprocal space and "tetrahedron" method [21] for the occupations. All geometries were fully optimized for total residual pressures on unit cells to within 0.5 kbar, and residual forces on atoms to within $10^{-3}$ mRy/a.u.

**Table 1**. GS and meta-stable (MS) states' properties in the OMC-$C_3^7$ method. The energies are in Ry/(formula unit) and are compared to the GS. Equilibrium lattice constants are in Angstrom, total magnetizations are in Bohr-magneton/(formula unit), and energy gap is in eV.

| State | initial occupations | $\Delta E$ | $a$ ($c$) | Tot. mag. | $E_{gap}$ |
|---|---|---|---|---|---|
| GS | [1110000],[0110010],[0110001] | 0.00000 | 5.4656 (5.4787) | 0.00 | 2.91 |
| MS1 | [1000011],[0100011],[0010011] | 0.00464 | 5.4888 (5.4327) | 0.00 | 2.22 |
| MS2 | [1001100],[0101100],[0011100],[0001110],[0001101] | 0.00838 | 5.4755 (5.4526) | 0.00 | 2.15 |
| MS3 | [1001010],[1001001],[0101010],[0011010],[0001011] | 0.05790 | 5.4354 (5.4628) | 0.00 | metal |
| MS4 | [1010010] | 0.06170 | 5.4479 (5.4543) | 0.00 | metal |
| MS5 | [1000110],[1000101],[0100110],[0010110] | 0.06203 | 5.4362 (5.4694) | 0.00 | metal |
| MS6 | [0111000],[1101000],[0101001],[0011001],[1011000] | 0.06206 | 5.4444 (5.4882) | 0.00 | metal |
| MS7 | [0100101],[1100100],[0010101],[1010100] | 0.06865 | 5.4588 (5.4571) | 0.00 | metal |
| MS8 | [0000111] | 0.06979 | 5.4446 (5.4418) | 0.00 | metal |
| MS9 | [0110100] | 0.08088 | 5.4703 (5.4248) | 0.00 | metal |
| MS10 | [1100001] | 0.09682 | 5.4151 (5.4954) | 0.00 | metal |
| MS11 | [1100010],[1010001] | 0.11800 | 5.4022 (5.4818) | 0.00 | metal |

### 2-A. OMC Method

In this method, the starting magnetization for oxygen atoms is set to zero values and for U atoms they are set to $\pm 0.5$ for making anti-ferromagnetic (AFM) configuration along the *z* direction.

Moreover, the initial values for the diagonal elements of occupation matrices in Hubbard correction are set to zero and unity ($n_\sigma = 0,1$ ; $\sigma = \uparrow, \downarrow$). Since in the present work we apply on-site Hubbard corrections to only 5*f* localized orbitals, the dimension of the occupation matrices are equal to 7. On the other hand, since in our employed U-atom pseudo-potential the 5*f* orbital is occupied by 3 electrons, then we can have $C_3^7 = 35$ different ways for occupying the diagonal elements of $7 \times 7$ matrix by 3 electrons: [1110000], [1101000], [1100100], … , [0001011], [0000111]. Among these occupation configurations the ones which lead to the lowest energy are taken as the initial occupations for GS. All others lead to metastable states. On the other hand, some researchers have argued that only two *f* orbitals of U atoms should be filled [10], and therefore ending up with a $C_2^7 = 21$ different ways for occupying the diagonal elements: [1100000], [1001000], [1000100], … , [0000101], [0000011]. We have considered both cases in our calculations.

## 2-B. SMC Method

In the SMC method [13], the starting magnetizations for the U atoms is set to $\pm 0.5$ as in the OMC method, but the starting magnetization for the O-atoms, $\varsigma$, are considered as new degrees of freedom. For the initial configurations of electrons, different values for starting magnetization in the interval $-1.0 \leq \varsigma \leq +1.0$ with step $\Delta\varsigma = 0.1$ were considered. This new degrees of freedom may be taken into account in two ways: (a)-All-equivalent O atoms, and (b)-Two inequivalent O atoms. In the former case, all oxygen atoms in the unit cell take the same values for $\varsigma$, and in the latter case, since there are two inequivalent uranium atoms with magnetizations $\pm 0.5$, we may consider two independent degrees of freedom for oxygen atoms which are denoted by $\varsigma_1$ and $\varsigma_2$. The situation is shown in Fig. 2. In this method the number of SCF calculations, 441, is much more than that in OMC, which equals at most 35.

## 3. Results and Discussions

### 3.A. OMC Results

**3.A.1.** $C_3^7$ **– occupations:** The self-consistent solutions in the OMC method with $C_3^7$ occupations resulted in 12 local minima, the lowest energy one is identified as the GS. The results are summarized in Table 1. It was already shown [10] that when the symmetries of wavefunctions were not included in the solutions of the KS equations, all metastable states turn into insulators with different gaps. In the case of OMC-$C_3^7$ method, we solved the KS equations without any symmetries for the wavefunctions other than the time reversal symmetry, and most of metallic metastable states turned into insulators. These calculations were done for cubic structure with experimental lattice constant of 5.47 Angstrom. The results are shown in Table 2. The states tabulated in Table 2 are named according to the configuration order and not the energy order.

**3.A.2.** $C_2^7$ – occupations: To test the validity of arguments [10] which led to $C_2^7$ configurations, we have performed the calculations and tabulated the results in Table 3. Comparison of Table 1 and Table 3 reveals that the GS and first two metastable states are the same in $C_3^7$ and $C_2^7$ schemes. However, the latter scheme predicts two more metastable states with gap. So, $C_2^7$ is not the only scheme to find the GS. In addition, we see that the predicted lattice constant for the GS, 5.4656 (5.4787) Angstrom is in excellent agreement with experimental value, 5.47 Angstrom. However, the predicted KS band gap, 2.91 eV, is about 0.7 eV larger than the experimental gap.

**3.B. SMC Results**

It was shown [13] that for the all-equivalent-O scheme, one obtains seven different classes of energetic and structural properties. Also, the calculated total magnetizations, $M_{tot} = \int_{cell} (n_\uparrow - n_\downarrow) d^3r$, had shown that the GS as well as the six meta-stable states had zero total magnetizations. The important result in the SMC method is that, in contrast to that of OMC method, for the GS the spin-up and spin-down electrons of O atoms are not symmetrically distributed. That is, the GS has a value $\varsigma \neq 0$. The situation is schematically shown in Fig. 3.

In the two-inequivalent-O scheme, we have optimized the structures for 441 different combinations of $\varsigma_1$ and $\varsigma_2$ values and obtained 16 local minima for the energies. The energy results are shown in Fig. 4. As is seen from Fig. 4, there are 16 local-minimum states and most of $\varsigma_1$ and $\varsigma_2$ combinations lead to the GS. This figure also shows that the properties are invariant under the transformation $(\varsigma_1, \varsigma_2) \rightarrow (-\varsigma_1, -\varsigma_2)$. More details (complementary to Fig. 4) on the result for two-inequivalent-O atoms scheme are presented in Table 4.

**Table 2**. Band gaps of metastable states for OMC-$C_3^7$ method when the symmetry of the wavefunctions were not included in the solutions of the KS equations. The states are sorted by the initial configurations of occupation matrices.

| State | initial occupations | $E_{gap}$ (eV) |
|---|---|---|
| GS* | [1110000] | 2.90 |
| MS*1 | [1101000] | 2.19 |
| MS*2 | [1100100] | 2.51 |
| MS*3 | [1100010] | 2.83 |
| MS*4 | [1100001] | 1.89 |
| MS*5 | [1011000] | 2.58 |
| MS*6 | [1010100] | 2.51 |
| MS*7 | [1010010] | 1.89 |
| MS*8 | [1010001] | 2.79 |
| MS*9 | [1001100] | 2.15 |
| MS*10 | [1001010] | 2.36 |
| MS*11 | [1001001] | 2.36 |
| MS*12 | [1000110] | 2.38 |
| MS*13 | [1000101] | 2.36 |
| MS*14 | [1000011] | 2.21 |
| MS*15 | [0111000] | not conv. |
| MS*16 | [0110100] | not conv. |
| MS*17 | [0110010] | 2.84 |
| MS*18 | [0110001] | 2.84 |
| MS*19 | [0101100] | 2.57 |
| MS*20 | [0101010] | 2.57 |
| MS*21 | [0101001] | 2.21 |
| MS*22 | [0100110] | 2.82 |
| MS*23 | [0100101] | not conv. |
| MS*24 | [0100011] | metal |
| MS*25 | [0011100] | 2.57 |
| MS*26 | [0011010] | 2.21 |
| MS*27 | [0011001] | 2.57 |
| MS*28 | [0010110] | not conv. |
| MS*29 | [0010101] | 2.86 |
| MS*30 | [0010011] | metal |
| MS*31 | [0001110] | 2.35 |
| MS*32 | [0001101] | 2.35 |
| MS*33 | [0001011] | metal |
| MS*34 | [0000111] | not conv. |

**Table 3**. The same as in Table 1 but for the OMC-$C_2^7$ method.

| State | initial occupations | $\Delta E$ | a (c) | Tot. mag. | $E_{gap}$ |
|---|---|---|---|---|---|
| GS | [110000] | 0.00000 | 5.4656 (5.4787) | 0.00 | 2.91 |
| MS1 | [1000001],[0000011] | 0.00464 | 5.4888 (5.4327) | 0.00 | 2.22 |
| MS2 | [0001100] | 0.00838 | 5.4754 (5.4523) | 0.00 | 2.15 |
| MS3 | [0010010] | 0.02426 | 5.5068 (5.3983) | 0.00 | metal |
| MS4 | [0001010],[0001001] | 0.05790 | 5.4354 (5.4628) | 0.00 | metal |
| MS5 | [0000110],[0000101] | 0.06203 | 5.4362 (5.4694) | 0.00 | metal |
| MS6 | [0101000],[0011000] | 0.06206 | 5.4444 (5.4882) | 0.00 | metal |
| MS7 | [1000100] | 0.06370 | 5.5209 (5.5176) | 0.00 | 1.00 |
| MS8 | [1001000] | 0.06608 | 5.5265 (5.5020) | 0.00 | 0.82 |
| MS9 | [0100100],[0010100] | 0.06865 | 5.4588 (5.4571) | 0.00 | metal |
| MS10 | [0100001] | 0.09682 | 5.4151 (5.4954) | 0.00 | metal |
| MS11 | [0100010],[0010001] | 0.11800 | 5.4022 (5.4818) | 0.00 | metal |
| MS12 | [1100000],[1010000] | 0.13873 | 5.4992 (5.5028) | 0.00 | metal |

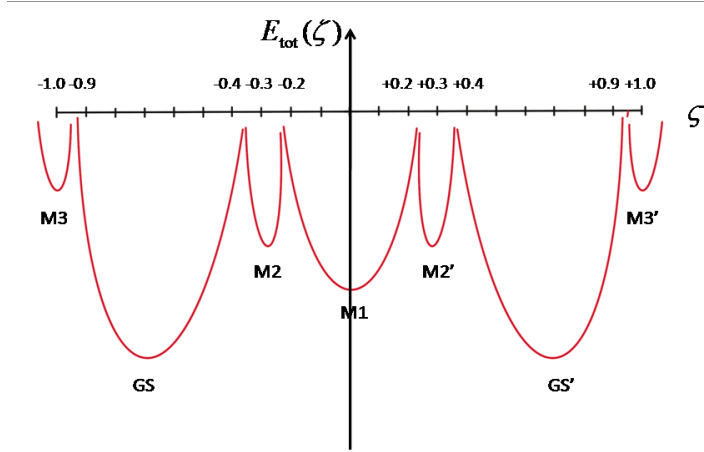

**Fig. 3.** Schematic plot of local minima and their corresponding starting magnetizations. The depths of the minima (in arbitrary units) are consistent with their energy values. As is seen, the GS and GS' cover the largest interval, but away from the zero starting magnetization.

Here, we have also performed SCF calculation with no symmetry (other than the time-reversal) for the wavefunctions, and did not find any significant differences in the number of "insulator" metastable states, and therefore do not report the results.

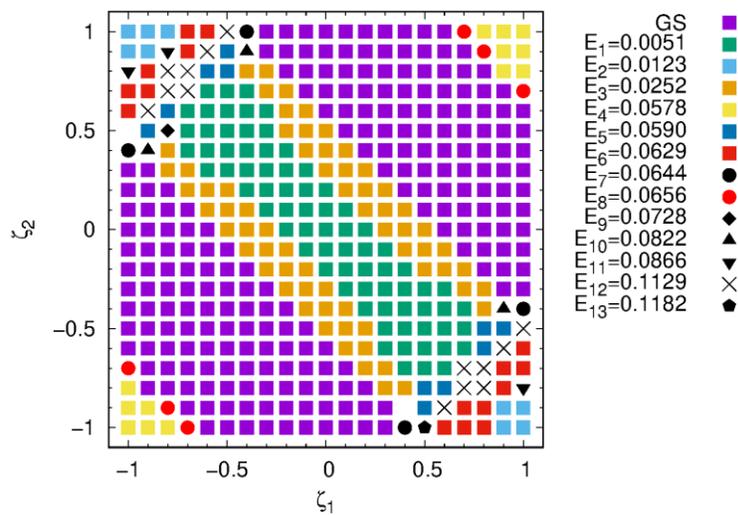

**Fig. 4.** Local-minimum states in two-inequivalent-O scheme. It is seen that for the GS, the values of $\varsigma_1$ and $\varsigma_2$ are not simultaneously zero and most of $\varsigma_1$ and $\varsigma_2$ combinations lead to the GS than the other 15 metastable states. Also invariance under $(\varsigma_1, \varsigma_2) \rightarrow (-\varsigma_1, -\varsigma_2)$ transformation is clearly seen.

Finally, we address the most important differences of the ground state properties in the two schemes of OMC and SMC. Firstly, the GS energy obtained in SMC is slightly higher than that of OMC by 0.0022 Ry/(formula unit). Secondly, the KS gap in SMC-GS, 2.18 eV, is very close to the experimental value compared to that of OMC-GS, 2.91 eV; and thirdly, as is seen from the plots of density of states (DOS) in Fig. 5, in the GS of OMC scheme the up-spin and down-spin electrons are symmetrically distributed over the states while in the SMC scheme even though the total magnetization vanishes but the symmetry between up and down spins is broken.

**Table 4.** GS and meta-stable (MS) states' properties in the SMC method with two-inequivalent-O atoms. The energies are in Ry/(formula unit) and are compared to the GS. Equilibrium lattice constants are in Angstrom, total magnetizations are in Bohr-magneton/(formula unit), and energy gap is in eV. Because of $(\varsigma_1,\varsigma_2) \rightarrow (-\varsigma_1,-\varsigma_2)$ symmetry in properties, they are not brought here.

| State | $\{(\varsigma_1, \varsigma_2)\}$ | $\Delta E$ | $a$ ($c$) | Tot. mag. | $E_{gap}$ |
|---|---|---|---|---|---|
| GS | (-1.0, [-0.7,+0.2]), (-0.9, [-0.8,+0.3]), (-0.8, [-0.9,+0.3]), (-0.7, [-1.0,+0.1]), (-0.6, [-1.0,0.0]), (-0.5, [-1.0,-0.1]), (-0.4, [-1.0,-0.2]), (-0.3, [-1.0,-0.3]), (-0.3, [+0.8,+0.9]), (-0.2, [-1.0,-0.4]), (-0.2, [+0.8,+1.0]), (-0.1, [-1.0,-0.5]), (-0.1, [+0.7,+1.0]), (0.0, [-1.0,-0.6]), (0.0, [+0.6,+1.0]) | 0.00000 | 5.4765 (5.4566) | 0.00 | 2.18 |
| MS1 | (-0.7, [+0.4,+0.6]), (-0.6, [+0.3,+0.7]), (-0.5, [+0.2,+0.7]), (-0.4, [+0.2,+0.7]), (-0.3, [+0.1,+0.6]), (-0.2, [0.0,+0.5]), (-0.1, [-0.1,+0.4]), (0.0, [-0.2,+0.2]) | 0.00240 | 5.4888 (5.4327) | 0.00 | 2.22 |
| MS2 | (-1.0, [+0.9,+1.0]), (-0.9, [+0.9,+1.0]), (-0.8,+1.0) | 0.00614 | 5.4755 (5.4526) | 0.00 | 2.15 |
| MS3 | (-0.8,+0.4), (-0.7, [+0.2,+0.3]), (-0.6, [+0.1,+0.2]), (-0.5, [0.0,+0.1]), (-0.4, [-0.1,+0.1]), (-0.4,+0.8), (-0.3, [-0.2,0.0]), (-0.3, +0.7), (-0.2, [-0.3,-0.1]), (-0.2, [+0.6,+0.7]), (-0.1, [-0.4,-0.2]), (-0.1, [+0.5,+0.6]), (0.0, [-0.5,-0.3]), (0.0, [+0.3,+0.5]) | 0.01218 | 5.4981 (5.4150) | 0.00 | 1.90 |
| MS4 | (-1.0, [-0.9,-0.8]), (-0.9, -1.0), (-0.8, -1.0) | 0.02848 | 5.4504 (5.4746) | 0.00 | metal |
| MS5 | (-0.8, +0.5), (-0.5, +0.8) | 0.029225 | 5.4621 (5.4485) | 0.00 | metal |
| MS6 | (-1.0, [+0.6,+0.7]), (-0.9, +0.8), (-0.7, +1.0), (-0.6, +1.0) | 0.031185 | 5.4562 (5.4578) | 0.00 | metal |
| MS7 | (-1.0, +0.3) | 0.031215 | 5.4719 (5.4516) | 0.00 | metal |
| MS8 | (-0.9, +0.4), (-0.4, +0.9) | 0.031775 | 5.4533 (5.4720) | 0.00 | metal |
| MS9 | (-0.9, -0.9) | 0.032190 | 5.4552 (5.4590) | 0.00 | metal |
| MS10 | (-0.3, +1.0) | 0.038175 | 5.4705 (5.4507) | 0.00 | metal |
| MS11 | (-1.0, +0.8) | 0.042295 | 5.4559 (5.4330) | 0.00 | metal |
| MS12 | (-0.9, [+0.5,+0.7]), (-0.8, [+0.6,+0.8]), (-0.7, [+0.7,+0.9]), (-0.6, [+0.8,+0.9]), (-0.5, +0.9) | 0.055675 | 5.4354 (5.4628) | 0.00 | metal |
| MS13 | (-1.0, [+0.4,+0.5]), (-0.5, +1.0), (-0.4, +1.0) | 0.058130 | 5.4257 (5.4870) | 0.00 | metal |
| MS14 | (-0.8, +0.9) | 0.064050 | 5.4390 (5.4467) | 0.00 | metal |
| MS15 | (-1.0, -1.0) | 0.066290 | 5.4315 (5.5120) | 0.00 | metal |

### 3.C. Searching for Global Minimum over Larger Subspaces

The differences between OMC-GS and SMC-GS imply that the two methods span different subspaces of electron densities and one may do searching for global minimum over larger subspaces. Here, we have considered two different extended subspaces: i)- applying OMC independently for up-spin and down-spin uranium atoms, ii)- applying OMC for each starting magnetization of O atoms.

### 3. C. 1. Independent OMC for U1 and U2 atoms

In this method, we have solved the KS equations using all $C_2^7 \times C_2^7 = 441$ different initial occupation matrices and obtained 80 different metastable states closely lying within 0.2467 Ry energy interval. The GS was obtained for two different combinations of occupations for up-spin and down-spin uranium atoms, respectively: ([0110000], [0110000]) and ([0100001], [0110000]), which is energetically identical with the symmetric OMC result explained above.

### 3. C. 2. OMC on top of SMC

Here, we apply the OMC method on U atoms for each value of $0 \leq \varsigma \leq 1$ with step $\Delta\varsigma = 0.1$ for O atoms. The results show that the occupation matrices leading to the GS are not independent from the starting magnetizations of O atoms. The results are presented in Table 5.

**Table 5.** Initial occupations of diagonal elements of *f* Hubbard orbitals that lead to GS for the corresponding given starting magnetization of O atoms (column one). Third and fourth columns specify the number of corresponding metastable states and the range, in Ry, over which they are distributed.

| starting mag. $\varsigma$ | GS initial occupations | number of MS states | energy range (Ry) |
|---|---|---|---|
| 0.0 | [1110000],[0110100],[0110010],[0110001] | 16 | 0.21922 |
| 0.1 | [1110000],[0110100],[0110010],[0110001] | 13 | 0.21922 |
| 0.2 | [1110000],[1100010],[1100001],[0110010] | 20 | 0.21922 |
| 0.3 | [1100010],[1100001] | 22 | 0.16105 |
| 0.4 | [1100010],[1100001] | 19 | 0.14337 |
| 0.5 | [1100010],[1100001] | 23 | 0.15528 |
| 0.6 | [1100010],[1100001] | 21 | 0.13946 |
| 0.7 | [1100010] | 19 | 0.12196 |
| 0.8 | [1100010],[1100001] | 25 | 0.13387 |
| 0.9 | [1100010],[1100001] | 18 | 0.13930 |
| 1.0 | [1110000],[1101000],[1100100],[1100010],[1100001] | 17 | 0.16590 |

As is seen from Table 5, depending on the initial magnetization of oxygen atoms, the diagonal elements of the initial occupation matrices for GS varies. This result crucially shows that ignoring the starting magnetization for oxygen atoms in OMC method may result in a metastable state. Moreover, as is seen from third and fourth columns of Table 5, the number of metastable states and well as the range of energy that they are spread out, depends on the starting magnetization of O atoms. Finally, whenever no symmetry applies to the cell (because of a defect for example), then to find the best candidate for the GS one has to consider occupation matrices for all Hubbard atoms independently.

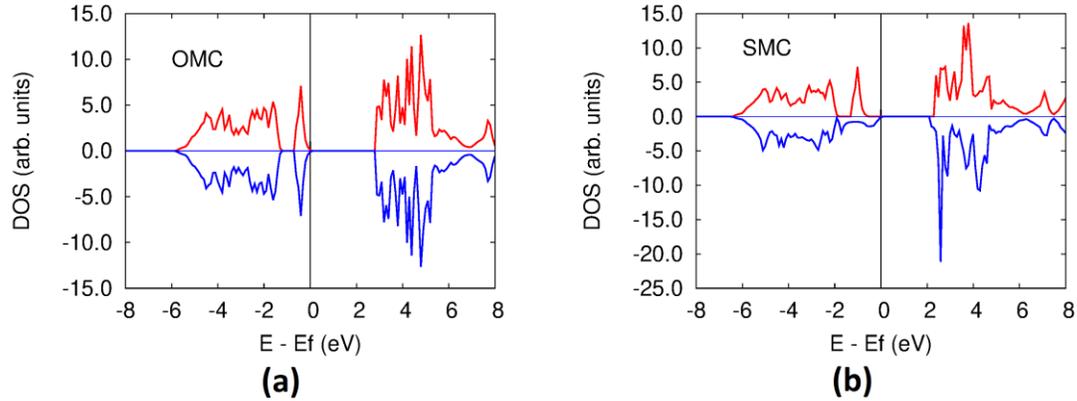

**Fig. 5.** Density of states (DOS) for spin-up (purple color) and spin-down (green color) for (a)- OMC and (b)- SMC methods. In OMC it is symmetric but in SMC the symmetry is broken.

## 4. Conclusions

To overcome the incorrect prediction of metallic behavior for $UO_2$ by ordinary DFT method, one may use a Hubbard correction for the energy contributions of localized orbitals, which is called the DFT+U method. However, the application of DFT+U to determine the GS properties is somewhat tricky in that one should take care not to stuck in higher-energy metastable states which may have very different structural and electronic properties. The occupation-matrix-control (OMC) method was one way to search for the global minimum among lots of local minimum states. In the OMC method, people constrain the starting magnetization of O atoms to be zero and use different possible initial occupation matrices for the localized orbital. However, in the SMC method we release the used constraint in the OMC method and let the starting magnetizations of O atoms vary between -1.0 and +1.0. It turns out that in this way one reproduces different sets of metastable states. It was shown that in the SMC method one obtains more metastable states than that of OMC. Comparing the GS properties in the two method showed very similar lattice constants and total magnetizations. However, even though the KS gap in SMC-GS, 2.18 eV, is very close to the experimental value compared to that of OMC-GS, 2.91 eV, the GS energy obtained in SMC is slightly higher than that of OMC by 0.0022 Ry/(formula unit). The differences between the two ground states imply that the two methods span different subspaces of electron densities and one may do searching for global minimum over larger subspaces in which both initial magnetization of O atoms and the occupation matrices of U atoms are involved. Finally, for a unitcell with no symmetry, to find the best candidate for the GS, one has to deal with the occupation matrices of all Hubbard atoms independently.

## Acknowledgements

This work is part of research program in School of Physics and Accelerators, NSTRI, AEOI.

## Data Availability

The raw or processed data required to reproduce these results can be shared with anybody interested upon sending an email to M. Payami.